\documentclass[12pt,letterpaper]{article}
\usepackage{amsmath,amssymb}
\usepackage{graphicx,color}
\usepackage[breaklinks]{hyperref}
\pdfoutput=1
\usepackage[bf]{caption}

\newcommand{\rr}{\mathbb{R}}

\def\theequation{\arabic{section}.\arabic{equation}}

\newcommand{\be}{\begin{equation}}
\newcommand{\ee}{\end{equation}}
\newcommand{\ba}{\begin{aligned}}
\newcommand{\ea}{\end{aligned}}
\newcommand{\ben}{\begin{displaymath}}
\newcommand{\een}{\end{displaymath}}
\newcommand{\bea}{\begin{eqnarray}}
\newcommand{\eea}{\end{eqnarray}}
\newcommand{\bean}{\begin{eqnarray*}}
\newcommand{\eean}{\end{eqnarray*}}

\newcommand{\p}{\partial}

\usepackage{amsmath,amssymb}
\usepackage{graphicx,color}

\def\th {\theta}
\def\a {\alpha}

\def\d {\delta}
\def\e {\epsilon}
\def\s {\sigma}
\def\e {\epsilon}

\def\m{\mu}
\def\n{\nu}

\def\o{\omega}

\definecolor{green}{rgb}{0,0.5,0}

\def\p{\partial}

\long\def\symbolfootnote[#1]#2{\begingroup
\def\thefootnote{\fnsymbol{footnote}}\footnote[#1]{#2}\endgroup}

\begin{document}

\begin{titlepage}
\vspace{10pt} \hfill {HU-EP-10/76} \vspace{20mm}
\begin{center}

{\Large \bf Coordinate representation of particle dynamics in AdS\\[2mm]
 and in generic static spacetimes  }


\vspace{45pt}

{Harald Dorn,$^a~$
George Jorjadze,$^{a,\,b}~$
Chrysostomos Kalousios,$^a$ Jan Plefka $^a$
\symbolfootnote[2]{\tt{\{dorn,jorj,ckalousi,jan.plefka\}@physik.hu-berlin.de}}
}
\\[15mm]

{\it\ ${}^a$Institut f\"ur Physik der
Humboldt-Universit\"at zu Berlin,}\\
{\it Newtonstra{\ss}e 15, D-12489 Berlin, Germany}\\[3mm]
{\it${}^b$Razmadze Mathematical Institute,}\\
{\it M. Aleksidze 1, 0193, Tbilisi, Georgia}

\vspace{20pt}

\end{center}

\vspace{40pt}

\centerline{{\bf{Abstract}}}
\vspace*{5mm}
\noindent
We discuss  the quantum dynamics of a  particle in static curved spacetimes 
in   a  coordinate representation. The scheme is based on the analysis of the squared energy operator $E^2,$ which is quadratic in momenta and contains a scalar curvature term. Our main emphasis is on  AdS spaces, where this term is fixed by the isometry group. As a byproduct the isometry generators are constructed and the energy spectrum is reproduced. In the massless case the conformal symmetry is realized as well. We show the equivalence between this quantization and the covariant quantization, based on the Klein-Gordon type equation in AdS. 
We further demonstrate that the two quantization methods in an arbitrary 
$(N+1)$-dimensional static spacetime are equivalent to each other if the scalar curvature terms both in the operator $E^2$ and in the Klein-Gordon type equation have the same coefficient equal to $\frac{N-1}{4N}$.

\vspace{15pt}
\end{titlepage}

\newpage

\tableofcontents


\section{Introduction}

In generic spacetimes the notion of a quantized particle becomes dependent
on the local coordinate system \cite{B-D}. Only in the presence of a high degree of symmetry one
is able to define particles in a global covariant manner. In Anti-de Sitter 
space AdS$_{N+1}$, like in the most prominent case of  Minkowski
space $\mathbb{R}^{1,N}$,  particles 
are identified with unitary irreducible representations of the corresponding 
isometry groups \cite{fronsdal}. 

The aim of this paper is twofold. On one side we want to make a contribution
to the study of the quantization problem for a relativistic particle in a
generic static spacetime. There, at least, the spectral problem for the energy
in certain suitably chosen coordinate systems is a reasonable one. Its study
includes ordering problems. We will show that even in the absence of 
further symmetries, one nevertheless can fix the ordering ambiguity. For this
purpose we insist
on general covariance with respect to space and require equivalence
of the quantization based on the Klein-Gordon equation
to quantization after Hamiltonian reduction realized in a coordinate 
representation.

The second aim is the application of the coordinate representation, to be developed
in the body of the paper, to the quantized particle in  AdS$_{N+1}$. In this
case the ordering ambiguity can be fixed exclusively within the coordinate 
representation by demanding that the algebra of the isometry generators $\mathfrak{so}(2,N)$ can be realized. In this way we will find a coordinate version of the
unitary irreducible representations of SO$(2,N)$. Of course we will reproduce
all the well-known results concerning the spectrum of the relevant quantum numbers,
see e.g. \cite{ads-review}. With the AdS/CFT correspondence in mind
we present this technique also because of its potential use for string quantization
in AdS$_{N+1}$.

The paper is organized as follows. We first start with some introductory aspects 
of the coordinate representation in generic static spacetimes and then specialize 
to  AdS$_{N+1}$ with its  SO$(2,N)$ representation. After this, still for AdS$_{N+1}$,
a comparison with the spacetime covariant treatment based on the 
Klein-Gordon equation is made. Finally we come back to generic static
spacetimes.

\vspace{3mm}

Let us consider a $(N+1)$-dimensional   spacetime with coordinates $x^\mu,$ $\mu=(0,1,...,N)$ and a static metric tensor
\be\label{metric g}
g_{\mu\nu}=\left(\begin{array}{cc}
              g_{00}(x) & 0 \\
              0 & g_{mn}(x)
            \end{array}\right)~,
\ee
where $g_{00}=-e^{f}$ and $g_{mn}$ are functions only of the spatial coordinates $x^n$ ($n=1,...,N$).

The dynamics of a particle in this background is described by the action
\be\label{action g}
S=\int \mbox{d}\tau \Big(p_\m\,\dot x^\m+\lambda\left(g^{\m\n}p_\m p_\n+M^2\right)\Big)~,
\ee
and in the gauge $x^0=\tau$ it reduces to an ordinary Hamiltonian system
\be\label{action h}
S=\int \mbox{d}\tau \Big(p_n\,\dot x^n-E(p, x)\Big)~,
\ee
where $E(p,x)=-p_0>0$ is the particle energy.

From the mass shell condition $g^{\m\n}p_\m p_\n+M^2=0$ one obtains the squared energy  
\be\label{E^2 g}
E^2=e^{f(x)}\,g^{mn}(x)~p_mp_n+M^2\,e^{f(x)}~.
\ee
This function can be associated with the Hamiltonian of a non-relativistic particle moving in the potential 
$M^2\,e^{f(x)}$ in a curved background with the metric tensor  
\be\label{x metric}
h_{mn}(x)=e^{-f(x)}\,g_{mn}(x)~.
\ee

Quantizing this system in the coordinate representation, one gets the Hilbert space with wave functions $\psi(x)$ and the scalar product
 \be\label{scalar product g}
\langle \psi_2 | \psi_1 \rangle =  \int \mbox{d}^Nx \,\sqrt{h(x)} ~ \psi_2^*(x) \, \psi_1(x)~,
\ee\,
where $h(x)=\mbox{det}\,h_{mn}(x)$. The Hermitian momentum operators are $p_n=-i\p_n-\frac{i}{4}\p_n\log h$.

Since (\ref{E^2 g}) is quadratic in momenta, the ordering ambiguities contain at
most second derivatives of $h_{mn}(x)$. Requiring general covariance in the reduced manifold, the freedom due to ordering ambiguities is parameterized by a constant 
$a$ in  
\be\label{E^2 g op}
E^2=-\Delta_h+a\mathcal{R}_h(x)+M^2\,e^{f(x)}~,
\ee
where  $\Delta_h$ is the covariant Laplace operator for the metric tensor $h_{mn},\,$ and 
$\mathcal{R}_h$ denotes the corresponding scalar curvature \cite{DeWitt-R} (for further references see \cite{basti}). One of our tasks is to fix the value of $a$.

\setcounter{equation}{0}

\section{Classical description of AdS particle}

\noindent
$\mbox{AdS}_{N+1}$ is realized as a hyperboloid $X^AX_A=-R^2$, where $X^A,$ $\,A=(0',0,1,...,N)$ are coordinates of the embedding space $\rr^{2,\,N}$, which we parameterize by\footnote{For notational  convenience we write the spatial coordinates with down indices.}
\be\label{coordinates}
X^{0'}=\frac{R\,\sin\th}{\sqrt{1-x^2}}~,\qquad X^{0}=\frac{R\,\cos\th}{\sqrt{1-x^2}}~,\qquad X^n = \frac{R\,x_n}{\sqrt{1-x^2}}~,
\ee
with $x^2:=x_n x_n<1$. The polar angle $\th$ is interpreted as a dimensionless  time coordinate $\th=x^0$, since the induced metric tensor on the hyperboloid  has the structure \eqref{metric g} with
\be\label{induced metric}
g_{00}=-\frac{R^2}{1-x^2}~,\qquad  g_{mn}=\frac{R^2}{1-x^2}\left(\d_{mn}+\frac{x_m\,x_n}{1-x^2}\right)~.
\ee

If a vector field ${\cal V}={\cal V}^\m\p_\m$ generates spacetime isometry transformations, i.e. ${\cal L}_{\cal V}\,g_{\m\n}=0,\,$ then $J={\cal V}^\m\,p_\mu$ is a Noether integral of the system \eqref{action g}, and in the gauge $x^0=\tau$ it becomes $J={\cal V}^n(\tau,\,x)\,p_n-{\cal V}^0(\tau,\,x)\,E(p,x),$ where $E$ is the particle energy  obtained from \eqref{E^2 g}.

The isometries of $\mbox{AdS}_{N+1}$ are the $\mbox{SO}(2,N)$ transformations  generated by the vector fields ${\cal V}_{AB}$, whose action on the embedding space coordinates is   ${\cal V}_{AB}(X^C)=\d_A^C\,X_B-\d_B^C\,X_A.$ By \eqref{coordinates} we then find
\be\ba\label{V_AB}
&{\cal V}_{0'0}=-\p_\th~, &{\cal V}_{mn}=x_n\p_m-x_m\p_n~,~~~~~~~~~\\[2mm]
&{\cal V}_{n0'}=-x_n\cos\th\,\p_\th-\sin\th\,\,V_n~, \qquad &{\cal V}_{n0}=x_n\sin\th\,\p_\th-\cos\th\,\,V_n~,
\ea\ee
where $V_n=\p_n-x_nD$ and $D$ is the dilatation operator $D=x_m\,\p_m$.
The corresponding Noether  integrals 
are given by (again in the gauge $\tau =x^0=\theta $) 
\be\label{dyn integrals 2}
J_{0'0}=E=\sqrt{p^2-(p\cdot x)^2+\frac{M^2R^2  }{1-x^2}}~,\quad J_{mn}=p_m\,x_n-p_n\,x_m~,
\ee
\be\label{J0n}
J_{n\,0'}=E\,x_n\,\cos\tau+\big((p\cdot x)x_n-p_n\big)\sin\tau~,~~ J_{n\,0}=\big((p\cdot x)x_n-p_n\big) \,\cos\tau -E\,x_n\sin\tau~.
\ee

Note that the Casimir invariant of these integrals is constant $\frac{1}{2}\,J_{AB}\,J^{AB}=M^2R^2  $
and the minimal value of energy $E_0$  is equal to\footnote{More precisely,
$E$ is the particle energy measured in units of $1/R$.  } $E_0~=~MR.$

At $\tau=0$, the boost generators reduce   to
\be\label{J0n=}
J_{n\,0'}=E\,x_n~,\qquad \quad J_{n\,0}=(p\cdot x)x_n-p_n~.
\ee
We also use their complex combinations
\be\label{Zn=}
Z_{n}=J_{n\,0'}-iJ_{n\,0}~, \qquad \qquad Z_{n}^*=J_{n\,0'}+iJ_{n\,0}~,
\ee
which in quantum theory become lowering and raising operators for the energy spectrum.

The Poisson brackets of the functions \eqref{dyn integrals 2}, \eqref{J0n=} form the $\mathfrak{so}(2,N)$ algebra. Its compact part $\mathfrak{so}(2)\oplus\mathfrak{so}(N)$ is trivially realized by $E$ and the rotation generators $J_{mn},$ and the rest of the algebra can be written as
\bea\label{so(2,N) algebra 1}
\{E,\, Z_n\}=-iZ_n~,\quad \{Z_m,\, Z_n\}=0~,\quad\{Z_m,\, Z_n^*\}=2i\d_{mn}\,E-2J_{mn}~,\\
\{J_{lm},\, Z_n\}=\d_{ln}\,Z_m-\,\d_{mn}\,Z_l~.~~~~~~~~~~~~~~~~~~~~~~~~~~\label{so(2,N) algebra 2}
\eea

The squared energy  in AdS is given by
\be\label{E^2}
E^2=h^{mn}(x)\,p_m p_n+\frac{M^2R^2  }{1-x^2}~,
\ee
with   $h^{mn}(x)=\d_{mn}-x_m\,x_n.$ Its inverse is the matrix
\be\label{u metric}
h_{mn}(x)=\d_{mn}+\frac{x_m\,x_n}{1-x^2}~, \qquad \mbox{and  } \qquad  h=\frac{1}{1-x^2}~ .  
\ee
It is just  the metric tensor 
of the $N$-dimensional unit semi-sphere.

Concluding this section we consider the massless particle in AdS.
For $M=0$ the action \eqref{action g} is invariant under conformal transformations. In $\mbox{AdS}_{N+1}$ those, which are not isometries,  are generated by the conformal  Killing vectors ${\cal V}_A$ with the property (see for example \cite{DJ1})\footnote{As usual, the case $\mbox{AdS}_2$ is special and has to be treated separately. }  
\be\label{C_A}
{\cal V}_A(X^B)=R\,\d_A^B+\frac{X_A\, X^B}{R}\,~, \qquad\qquad {\cal L}_{{\cal V}_A}\,\,g_{\m\n}=\frac{2X_A}{R}\,g_{\m\n}~. 
\ee
Then,    from \eqref{coordinates} one gets
\be\ba\label{V_A}
{\cal V}_{0'}=\sqrt{1-x^2}\,(\cos\th\,\p_\th-\sin\th\,D)~, \quad {\cal V}_{0}=-\sqrt{1-x^2}\,(\sin\th\,\p_\th+\cos\th\,D)~,\\[2mm]
\quad {\cal V}_{n}=\sqrt{1-x^2}\,\p_n~.~~~~~~~~~~~~~~~~~~~~~~~~~~~~~~~~~~~~~~
\ea\ee
The corresponding dynamical integrals at $\tau=0$ become
\be\label{C_A=}
C_{0'}=-\sqrt{1-x^2}\,E~, \qquad C_0=-\sqrt{1-x^2}\,(p\cdot x)~, \qquad C_n=\sqrt{1-x^2}\,p_n~,
\ee
where $E$ is the energy of the massless particle, with
\be\label{E, M=0}
E^2=p^2-(p\cdot x)^2~.
\ee
This $E^2$  
coincides with the Hamiltonian of a free particle on the unit semi-sphere.
That is why the massless particle reaches  the boundary in a finite time. 

\setcounter{equation}{0}

\section{AdS particle in coordinate representation}

A consistent quantization of the particle dynamics in $\mbox{AdS}_{N+1}$ should provide a unitary irreducible representation of the $\mbox{SO}(2,N)$ group.
We use the classical expressions of dynamical integrals \eqref{dyn integrals 2}, \eqref{J0n=} and apply the coordinate representation. To specify the ordering prescription for the operators we first consider $E^2$. Since the scalar curvature of the unit sphere is equal to $N(N-1),$ the prescription \eqref{E^2 g op} provides the operator
\be\label{E^2 N}
E^2=-\sqrt{1-x^2}\,\,\p_m\,\,\frac{\d_{mn}-x_mx_n}{\sqrt{1-x^2}}\,\,\p_n  +a\,N(N-1)+\frac{M^2\,R^2}{1-x^2}~.
\ee
We will fix the parameter $a$ below. Since 
$E^2$ will turn out to be positive,   its positive square root defines the energy operator $E$.

The rotation generators in \eqref{dyn integrals 2} have no ordering ambiguity and they take the standard form
\be\label{Jmn}
J_{mn}=i (x_m\p_n-x_n\p_m)~.
\ee

For the ordering in  the boost generators $J_{n0'}$ \eqref{J0n=} we guess  
\be\label{boosts N 1}
J_{n0'}=\sqrt{E}~x_n\,\sqrt{E}~,
\ee
and get the commutator
\be\label{[E^2,Jn]}
[E^2,\,J_{n0'}]=\sqrt{E}\, \Big(N\,x_n-2V_n\Big)\,\sqrt{E}~.
\ee
On the other hand the $\mathfrak{so}(2,N)$ algebra requires  
\be\label{[E^2,Jn0']=}
[E^2,\,J_{n0'}]=2i J_{n0}\,E+J_{n0'}~ .  
\ee
Comparing these two commutators we obtain
\be\label{Jn0=}
J_{n0}=i \sqrt{E}~\left(V_n -\frac{N-1}{2}\,x_n\right)~\frac{1}{\sqrt{E}}~.
\ee
By \eqref{E^2 N}, \eqref{boosts N 1} and \eqref{Jn0=} we can verify the commutator similar to \eqref{[E^2,Jn0']=}
\be\label{[E^2,Jn0]=}
[E^2,\,J_{n0}]=-2i J_{n0'}\,E+J_{n0}~.
\ee
This condition is fulfilled if the constant term in $E^2$ is equal to $\frac{(N-1)^2}{4}.$ That fixes the coefficient of the scalar curvature
\be\label{a}
a=\frac{N-1}{4N}~.
\ee

Having fixed the ordering ambiguities by requiring the preservation of part of the consequences of the $\mathfrak{so}(2,N)$ algebra, we now discuss the issue of energy eigenfunctions and spectrum. Our Hilbert space is given by the wave functions $\Psi (x)$  with the scalar product
\be\label{scalar-prod}
\langle \Psi_2\vert\Psi_1\rangle~ =~\int _{x^2 < 1}~\frac{d^Nx}{\sqrt{1-x^2}}~\Psi_2 ^*(x)\,\Psi_1 (x) ~.
\ee

The ground state has to be a $\,\mbox{SO}(N)\,$ scalar function, which is annihilated by the lowering operators $Z_n=J_{n0'}-iJ_{n0}.$ These conditions are equivalent to the equations
\be\label{gs eq. N}
\left (x_n \Big(E_0  -\frac{N-1}{2}\Big)+V_n\right)\,\Psi_{E_0}(x^2)=0  ~,
\ee
where $ E_0  $ denotes the energy of the ground state. This yields  the wave function
\be\label{vacuum N}
\Psi_{E_0}  \sim (1-x^2)^{\frac{E_0   }{2}-\frac{N-1}{4}}~.
\ee
Since \eqref{vacuum N} should be an   eigenfunction of the operator \eqref{E^2 N} with the eigenvalue $ E_0   ^2,$ we can  relate the mass  to the lowest energy value by
\be\label{a,b}
M^2\,R^2=\left(E_0   -\frac{N}{2}\right)^2-\frac{1}{4}~.
\ee

The finiteness of the norm of the vacuum wave function \eqref{vacuum N} reproduces   the well known unitarity bound  $ E_0  >\frac{N}{2}-1\,\,$ \cite{B-F}. From \eqref{a,b} then follows that the two values of $E_0  $
\be\label{b,alpha}
 E_0^{\pm}  =\frac N 2\pm \sqrt{M^2R^2+\frac 1 4}~,
\ee
correspond to the same $M^2$, if $\,-\frac 1 4\leq M^2\,R^2<\frac 3 4,$ and for $M^2\,R^2\geq \frac 3 4$ only $E_0 ^{+}  $ is admissible.

The action of the raising operators $Z_n^*$ on the vacuum state
creates higher level eigenfunctions.
However,   already on the second level the states $Z_m^*Z_n^*\Psi_{E_0}$ with $m\neq n$ are not orthogonal to each other. Lower dimensional cases are exceptional and they are presented in the Appendix.
To find orthogonal eigenfunctions in higher dimensional cases we use the Casimir operator of  the   rotation generators
$L^2=\frac{1}{2}\,J_{mn}J_{mn}$ and rewrite the operator  \eqref{E^2 N} with \eqref{a} and \eqref{a,b}  as
\be\label{E^2 =L^2}
E^2=-4z(1-z)\p^2_{z}+2\big((N+1)z-N\big)\p_{z}+\frac{1}{z}\,L^2+\frac{(N-1)^2}{4}+ \frac{\left( E_0  -\frac{N}{2}\right)^2-\frac{1}{4}}{1-z}~,
\ee
with $z=x^2$. We look for the eigenfunctions of these operator in the form $\Psi=F(z)\,Y_L(\Omega),$ where $Y_L(\Omega)$ is a spherical harmonic, which is an eigenfunction of the
operator $L^2$ with the eigenvalue $L(L+N-2)$. Then, the eigenvalue problem
\be\label{evp}
E^2\Psi_\o\,=\,\o^2\Psi_\o
\ee
is solved by
\be\label{eigenfunction F}
F(z)=z^{\frac{L}{2}}(1-z)^{\frac{E_0}{2}-\frac{N-1}{4}}\,\,_2F_1(a,b,c;z)~,
\ee
where $_2F_1(a,b,c;z)$ is the hypergeometric function with the parameters
\be\label{a,b,c}
a=\frac{1}{2}\left(E_0+L-\o\right)~, \quad b=\frac{1}{2}\left(E_0+L+\o\right)~,\quad c=\frac{N}{2}+L~.
\ee
The regularity condition at $z=1$ requires $a=-n$ and one gets the energy spectrum (in agreement with \cite{fronsdal,B-F,ads-review}) 
\be\label{spectrum}
\o_{n,\,L}= E_0  +L+2n~.
\ee
It coincides with the spectrum of the  $N$-dimensional   normal ordered and shifted (by $ E_0   $) harmonic oscillator Hamiltonian  $\hat H=a_n^*\,a_n+ E_0  $ \cite{Dorn:2005ja}.  

It remains to check the full set of   commutation relations between the symmetry generators.  
Nontrivial are only the commutators corresponding to \eqref{so(2,N) algebra 1}. First note that from  \eqref{[E^2,Jn0']=} and \eqref{[E^2,Jn0]=} follows the commutator
$[E^2,Z_n^*]=Z_n^*\,(1+2E)$, which is equivalent to the statement: if $\Psi_\o$ is an eigenstate of $E^2$ with the eigenvalue $\o^2$, then  $Z^*_n\Psi_\o$ is also an eigenstate of $E^2$ with the eigenvalue $(\o+1)^2$. This implies $[E,Z_n^*]=Z_n^*.$   Now, only the commutation relations between the boost operators need discussion.  

Calculating the commutator $[J_{m0'}\,,\,J_{n0'}],$ we rewrite it in the form
\be\label{[J_{m0'},J_{n0'}]}
[J_{m0'},J_{n0'}]=\sqrt{E}\big(x_m\,E\,x_n\,E-x_n\,E\,x_m\,E\big)\frac{1}{\sqrt{E}}~,
\ee
and use the relation 
\be\label{[E,x]}
E\,x_n\,E-x_n\,E^2=\frac{N-1}{2}\,x_n-V_n~,
\ee
which easily follows from the commutator $[E,J_{n0'}]=iJ_{n0}$.
With the help of \eqref{[E,x]}, we find that the operator in the parentheses in \eqref{[J_{m0'},J_{n0'}]} is $J_{mn}$. Since $[E,J_{mn}]=0$,
we can neglect  the operators $\sqrt{E}$ in \eqref{[J_{m0'},J_{n0'}]} and obtain $[J_{m0'}\,,\,J_{n0'}]=iJ_{mn}$.
The other commutation relations of the boost generators are derived in a similar way. In particular, from
$[E,J_{n0}]=-iJ_{n0'}$ follows
\be\label{[E,V]}
E\,V_n\,E-V_n\,E^2=\left(\frac{(N-1)^2}{4}-E^2\right)x_n-\frac{N-1}{2}\,V_n~,
\ee
which together with \eqref{[E,x]} is helpful to check the commutator $[J_{m0}\,,\,J_{n0'}]=i\d_{mn} E$. 

The calculation of the Casimir operator can be done also with the help of \eqref{[E,x]} and it yields $E_0(E_0-N)$. Note that the relation of the Casimir to the lowest energy value has been renormalized relative to the classical expression $E_0^2$.  

Finally, we discuss the massless case.
According to \eqref{a,b} it corresponds   to 
\be\label{massless}
E_0^\pm   =\frac{N\pm 1}{2}~,
\ee
with the ground state wave functions $\Psi_{E_0^-}\sim 1,$ $~\Psi_{E_0^+}\sim \sqrt{1-x^2}.$

To construct the generators of conformal transformations we use the functions \eqref{C_A=} and first introduce the operator $C_{0'}$ similarly to the boost generators \eqref{boosts N 1}. Then, $C_{0}$ and $C_n$ can be defined by the commutators of $C_{0'}$ with $E^2$ and $J_{n0'}$, respectively. As a result we find
\be\ba\label{C_A quant}
C_{0'}=-\sqrt{E}\,\,\sqrt{1-x^2}\,\,\sqrt{E}~, \qquad C_0=i\sqrt{E}\,\,\sqrt{1-x^2}\left(D+\frac{N-1}{2}\right)\frac{1}{\sqrt{E}}~,\\[2mm]
C_n=-i\sqrt{E}\,\,\sqrt{1-x^2}\,\,\p_n\,\,\frac{1}{\sqrt{E}}~.~~~~~~~~~~~~~~~~~~~~~~~~~~~
\ea\ee
These operators together with the isometry generators realize the commutation relations of the conformal group $\mbox{SO}(2,N+1)$.
Like for the isometry generators, the check goes as follows: one first realizes the $\mathfrak{sl}(2)$ algebra with $E$, $C_{0'},$ $C_{0}$
and then derives operator identities similar to \eqref{[E,x]}, \eqref{[E,V]}, which become helpful to verify the other
commutation relations.
Introducing the lowering and raising operators $Z=C_{0'}-iC_0\,$  and $\,Z^*=C_{0'}+iC_0$, one finds
\be\label{Z,Z*}
Z\Psi_{E_0^+}\sim \Psi_{E_0^-}~, \qquad\qquad  Z^*\Psi_{E_0^-}\sim \Psi_{E_0^+}~.
\ee
Therefore, the conformal symmetry is realized on the states constructed by the action of the creation operators
$(Z_n^*\,,\,Z^*)$ on both ground states $\Psi_{E_0^-}$ and  $\Psi_{E_0^+}\,$ \cite{DJ1}.

For $\mbox{AdS}_2\,$  note that $\,E_0^-=0$  and the operator $1/\sqrt{E}$ becomes singular. That also makes the discussion of the conformal symmetry in $\mbox{AdS}_2$ special.  

\setcounter{equation}{0}

\section{Covariant quantization in $\mbox{AdS}$}

Covariant quantization of the particle dynamics in  $\mbox{AdS}_{N+1}$ is based on the analysis of the Klein-Gordon type equation 
\be\label{K-G eq}
(\Box -{\cal M}^2 )\Phi =0~,
\ee
where $\Box$ is the covariant d'Alembertian in $\mbox{AdS}_{N+1}$. Eq. \eqref{K-G eq} is a quantum analog of the mass shell condition $g^{\m\n}p_\m p_\n+M^2=0$ and, in general, one has to make the replacement 
\be\label{replacement}
g^{\m\n}p_\m p_\n ~~\mapsto ~~-\Box -\tilde a\,{\cal R}_g~,
\ee
where ${\cal R}_g$ is the spacetime scalar curvature and $\tilde a$ is a constant.
Since ${\cal R}_g$ is constant for $\mbox{AdS}_{N+1}$  and equal to $-\frac{N(N+1)}{R^2}$, this ambiguity has been 
included in ${\cal M}^2$
\be\label{M-M}
{\cal M}^2~=~M^2-\tilde a ~\frac{N(N+1)}{R^2}~. 
\ee  
Our task is now to fix the ambiguity of the value of ${\cal M}^2$ by requiring exact correspondence to the previous treatment in coordinate representation. 

The Hilbert space of the covariant quantization is formed by the positive frequency solutions of the equation \eqref{K-G eq}
\be\label{Psi=}
\Phi_\o(\theta,x)=e^{-i\o\theta}\phi_\o(x)~, \qquad \qquad \o>0~,
\ee
with the standard scalar product
\be\label{scalar product 2}
\langle \Phi_2 | \Phi_1 \rangle = \frac i 2 \int_{\theta=\tau} \mbox{d}^Nx \,\sqrt{-g} \,g^{00}\,\,\left({\partial_\theta}\Phi_2^*\,\,\Phi_1-\Phi^*_2\, \, {\partial_\theta}\Phi_1\right)~.
\ee

Using again the coordinates $x^\m=(\theta, x_n),\,$ by \eqref{induced metric} we find
\be\label{metric det}
g=-\frac{R^{2N+2}}{(1-x^2)^{N+2}}~,\qquad  g^{mn}=\frac{1-x^2}{R^2}\left(\d_{mn}-{x_m\,x_n}\right)~,
\ee
and the equation for $\phi_\o(x)$ obtained from \eqref{K-G eq} and \eqref{Psi=} takes the form of an eigenvalue problem
\be\label{E^2 N+1}
-(1-x^2)^{N/2}\,\p_m\frac{\d_{mn}-x_m\,x_n}{(1-x^2)^{N/2}}\,\p_n\phi_\o(x)+\frac{{\cal M}^2R^2}{1-x^2}\,\,
\phi_\o(x)=\o^2\,\phi_\o(x)~.
\ee
If we multiply this equation by $(1-x^{2  })^{\frac{1-N}{4}}$ and make the replacements
\be\label{replacements}
{\cal M}^2R^2 \mapsto  E_0(E_0  -N)~,  \qquad \phi_\o(x) \mapsto (1-x^2)^{\frac{N-1}{4}}\,\Psi_\o(x)~,
\ee
we get just the eigenvalue problem \eqref{evp} for the operator \eqref{E^2 =L^2}.

To justify these replacements and find the exact correspondence between the two quantization methods we
compare the corresponding scalar products and the symmetry generators.

When two wave functions \eqref{Psi=} have equal $\o$, the scalar product \eqref{scalar product 2} reduces to
\be\label{scalar prod N+1}
\langle \Phi_2 | \Phi_1 \rangle =
R^{N-1}\int\frac{\mbox{d}^Nx}{(1-x^2)^{N/2}}~\phi_{\o,\,_2}^*(x)\,\o\,\phi_{\o,_1}(x)~.
\ee
The local integration measure here differs from $\sqrt{h(x)}$ just by the factor compensated in the rescaling \eqref{replacements}.
The exact correspondence between the wave functions then takes the form
\be\label{phi-psi E}
\Phi_\o(\th,x)=\frac{e^{-i\o\theta}}{\sqrt{R^{N-1}\,\o}}\,(1-x^2)^{\frac{N-1}{4}}\,\Psi_\o(x)~,
\ee
where $\o$ is an eigenvalue of the operator $E$. For generic superpositions of energy eigenfunctions the correspondence \eqref{phi-psi E} implies 
\bea\label{NW}
\phi(x)&=& \int _{y^2< 1}~\frac{d^Ny}{\sqrt{1-y^2}}~ K(x,y)~\Psi(y)~,\nonumber\\
K(x,y)&=&\sum_{L,l,n}~\frac{(1-x^2)^{\frac{N-1}{4}}}{\sqrt{R^{N-1}\,\omega_{n,L}}}~\Psi _{n,L,l}(x)~\Psi^*_{n,L,l}(y)~,
\eea
with $l$ a collective (angular momentum ) index counting the degeneracy of the
energy levels \eqref{spectrum}. The function $K(x,y)$ is the AdS analog of the
Newton-Wigner function \cite{N-W} in Minkowski space  and $\phi(x)=\Phi(0,x)$.  

The wave functions $\Phi(\th,x)$ are scalar fields under the isometry transformations. This implies that
the symmetry generators of the covariant quantization are given by  
\be\label{J_AB}
J_{AB}=-i\,{\cal V}_{AB}~,
\ee
where ${\cal V}_{AB}$ are the vector fields of the isometry transformations \eqref{V_AB}.
The relation between the Casimir operator of the generators \eqref{J_AB} and the covariant d'Alembertian
\be\label{Casimir cov}
\frac{1}{2}\, J_{AB}\,J^{AB}=R^2\,\,\Box~,
\ee
justifies the replacement ${\cal M}^2R^2\rightarrow E_0(E_0   -N)$ in \eqref{E^2 N+1}.  

Note that this expression for  ${\cal M}^2$ together with \eqref{a,b} and \eqref{M-M} fixes also $\tilde a=\frac{N-1}{4N}$.  The bound 
$M^2R^2\geq-\frac{1}{4}$ stated after eq. \eqref{b,alpha} then corresponds to the 
Breitenlohner-Freedman bound ${\cal M}^2R^2 \geq  -\frac{N^2}{4}$ \cite{B-F}.  

Let us now  consider the relation between the symmetry generators \eqref{J_AB} and their counterparts in the coordinate representation. From the correspondence \eqref{phi-psi E} we find that the operators \eqref{J_AB} are mapped just to the operators constructed in the previous section. The rule for this map is the following:
one has to replace the operator $i\p_\th$ by $E$, then set $\th=0$ in \eqref{J_AB}, \eqref{V_AB}  and then multiply the obtained operators by $\sqrt{E}\,(1-x^2)^{\frac{1-N}{4}}$ from left and by $(1-x^2)^{\frac{N-1}{4}}/\sqrt{E}$ from right.

A last comment concerns  the massless case. In the covariant approach it is described by the equation
\be\label{M=0}
\left(\Box+\frac{N^2-1}{4R^2}\,\right)\Phi=0~,
\ee
corresponding to $ E_0^{\pm}   =\frac{N\pm 1}{2}$.
This equation is invariant under conformal transformations of the wave function $\Phi$ that infinitesimaly can be written as
 \cite{Jackiw} (see \eqref{C_A})  
\be\label{Conf map of Phi}
\Phi \mapsto \Phi+\e^A\left({\cal V}_A+\frac{N-1}{2R}\,X_A\right)\Phi~,
\ee
where $X_A$ are the embedding coordinates and ${\cal V}_A$ are the vector fields of the conformal transformations
\eqref{V_A}. One can check that the above described correspondence between the operators of the covariant quantization and the coordinate representation matches the generators of the conformal transformations as well.

\setcounter{equation}{0}

\section{Particle in a static spacetime}

At the end we return to  the general case started with the operator \eqref{E^2 g op}.
Its eigenvalue problem is equivalent to the equation
\be\label{ev problem}
\Delta_h\,\psi_\o(x)+\Big(\omega^2-a\mathcal{R}_h(x)-M^2\,e^{f(x)}\Big)\psi_\o(x)=0~.
\ee
According to \eqref{x metric} $h_{mn}$ is a rescaling of   $\hat g_{mn}$, the spatial part of the metric tensor. Therefore,  the scalar curvature here can be written as
\be\label{rescale R N}
{\mathcal{R}}_h=
e^{f}\,\left(\mathcal{R}_{\hat g}+(N-1)\Delta_{\hat g} f-\frac{(N-1)(N-2)}{4}\,\,g^{mn}\,\p_m f\,\p_n f\right)~.
\ee
The  Laplace operators of the rescaled metrics are related by
\be\label{Delta-h}
\Delta_h=e^{f}\left(\Delta_{\hat g}-\Big(\frac{N}{2}-1\Big)\,\,g^{mn}\,\p_m f\,\p_n\right)~.
\ee

Now we consider the covariant quantization of the same system, based on the Klein-Gordon type equation
with the replacement rule \eqref{replacement}
\be\label{KG-g}
(\Box -\tilde a\,\mathcal{R}_g-\tilde M^2 )\Phi =0~.
\ee

The positive frequency solutions of these equation $\Phi=e^{-i\o x^0}\,\phi_\o(x)$ form a   Hilbert space with the scalar product \eqref{scalar product 2} 
and  the wave function $\phi_\o(x)$ satisfies the equation
\be\label{phi eq}
\sqrt{\frac{e^{f}}{\hat g}}\,\,\p_m\,\sqrt{e^{f}\, \hat g}~g^{mn}\,\p_n\phi_\o(x)+\left(\o^2-\tilde a\,e^f\,\mathcal{R}_g-\tilde M^2e^f\right)\phi_\o(x)=0~,
\ee
 that can also be treated as an eigenvalue problem. To relate it with \eqref{ev problem},
we introduce the map between the eigenfunctions
\be\label{phi-psi map}
\phi_\o(x)=\frac{e^{\frac{1-N}{4}\,f(x)}}{\sqrt{\o}}\, \psi_\o  (x)~,
\ee
compatible with the scalar products \eqref{scalar product g}, \eqref{scalar product 2} and \eqref{scalar prod N+1}.

Using the relation between the scalar curvatures $\mathcal{R}_g$ and $\mathcal{R}_{\hat g}$
\be\label{R_g}
{\mathcal{R}}_{g}=
\mathcal{R}_{\hat g}-\Delta_{\hat g} f-\frac{1}{2}\,\,g^{mn}\,\p_m f\,\p_n f~,
\ee
and eqs. \eqref{rescale R N}-\eqref{Delta-h}, we find that
\eqref{ev problem} and \eqref{phi eq} are equivalent to each other  for
\be\label{conditions}
\tilde M=M~, \qquad \qquad \tilde a=a=\frac{N-1}{4N}~.
\ee
This value of the coefficient $\tilde a$ corresponds to the Weyl invariance for $M=0$ ( see e.g. \cite{Jackiw}).

\setcounter{equation}{0}

\section{Conclusions}
We discussed the quantization of a 
particle in $(N+1)$-dimensional static spacetimes after Hamiltonian reduction 
within a coordinate representation.
The key point of this analysis is the observation, that the expression
for the squared energy looks like the Hamiltonian for a non-relativistic
particle in a curved $N$-dimensional space. This fact has been
used in the literature at several places, see e.g. \cite{Castagnino:1989vf,Wang:2009ay,Greenwood:2009gp}. However,  
the use  of this observation for fixing some ordering ambiguities and 
for a complete realization of symmetry algebras in terms of operators
acting on position space wave functions is a new facet of this issue.  
The description of a quantum particle in terms of position dependent wave functions seems to be most naturally. But in order to respect the relativistic principles, the generators for transformations involving time (energy, boosts) become non-local.

The ordering ambiguities concern the value of a factor in front of the scalar 
curvature which has been fixed to $a=\frac{N-1}{4N}$. Applying in different
situations other principles or normalization conditions, one generates other
values for $a$, compare e.g. the value  appearing in the setting of \cite{Gervais:1976ws,DeWitt,basti}. For the isometry algebra of AdS$_{N+1}$ we found a new representation
in terms of operators acting on functions depending on $N$-dimensional space
coordinates. To get this representation, we had to solve further ordering problems
for all the other generators besides the squared energy.    

Usually one notes  a
posteriori that the Klein-Gordon equation for ${\cal M}^2=M_{_W}^2=-\frac{N^2-1}{4R^2}$
exhibits conformal invariance \cite{Fronsdal:1975ac, Avis:1977yn} and therefore $M^2={\cal M}^2-M_{_W}^2$
should be called {\it the} squared mass. Our analysis sheds some different light on the interpretation of the mass parameter
${\cal M}^2$ in the  AdS$_{N+1}$ Klein-Gordon equation. We use no field theoretical
arguments. On the basis of pure particle quantum mechanics our $M^2$ is a priori the squared mass, and the shift to  ${\cal M}^2$ is generated by the quantum mechanical
ordering effect.

As a byproduct we constructed the AdS analog of the Minkowski space Newton-Wigner
wave functions \cite{N-W}. It would be interesting to perform the sum 
explicitly and to compare its analytic properties to the Minkowski case and recent discussion in de Sitter space \cite{barata}.
Note that a generalization of our quantization scheme to non-static spacetimes with a global time coordinate is straightforward.

Furthermore, one could try to apply  the techniques of our paper to the quantization of strings in AdS 
using a gauge in which the energy density along the string is constant.

\section*{\large Acknowledgments}

This work has been supported in part by Deutsche Forschungsgemeinschaft via SFB 647 and by VolkswagenStiftung.
G.J. was also supported by GNSF.

\appendix

\setcounter{equation}{0}

\def\theequation{A.\arabic{equation}}

\section{Low dimensional cases}

\subsection*{Eigenfunctions in $\mbox{AdS}_2$}

In $\mbox{AdS}_2$ the scalar curvature term in \eqref{E^2 N} vanishes and the operator \eqref{E^2 N} becomes
\be\label{E operator u}
E^2=-(1-x^2)\p_x^2+x\p_x+\frac{E_0(E_0-1)}{1-x^2}~, 
\ee
with $x=x_1\in (-1,1)$.
The ground state
\be\label{vacuum=}
\Psi_{E_0}  (x)\sim (1-x^2)^{\frac{E_0}{2}}~,
\ee
which is annihilated by the lowering operator $Z=J_{10'}-iJ_{10}$, is an eigenfunction of the operator
\eqref{E operator u} with the eigenvalue $E_0^2$.  
The action of the raising operator $Z^*=J_{10'}+iJ_{10}$ on the ground state creates the first exited state
$\Psi_{E_1} (x)\sim x (1-x^2)^{\frac{E_0}{2}}$,   which is an eigenfunction of the operator \eqref{E operator u}  with the eigenvalue $(E_0+1)^2.$  
Continuing this process step by step, with the notation $\psi_k=\Psi_{E_k}$ and $\a=E_0$,  one finds
\be\label{psi_n=}
\psi_{k+1}(x) \sim [(\a+k)x-(1-x^2)\partial_x]\psi_{k}(x)~,
\ee
and therefore the eigenfunctions can be written in the form
\be\label{psi_n=1}
\psi_k(x)=c_{k,\,\a} \,(1-x^2)^{\frac{\a}{2}}\, C_k^\a(x)~,
\ee
where $C_k^\a(x)$ is a polynomial of order $k$ and $c_{k,\a}$ denotes a normalization constant.
According to \eqref{psi_n=}, the functions $C_k^\a(x)$
satisfy the recursive relations
\be\label{recur-rel}
C_{k+1}^{\a} (x) =(2\a+k)\,x\,C_k^{\a} (x)-(1-x^2) \frac{\mbox{d}}{\mbox{d}x} C_k^{\a}(x)~,
\ee
of the Gegenbauer Polynomials. These functions are orthogonal to each other under the scalar
product with the weight $(1-x^2)^{\a-{1}/{2}}$  and the normalization coefficients are given by
\be\label{coefficient}
c_{k,\,\a} = 2^\a\, \Gamma(\a)\left[ \frac{(k+\a  )k!}{2\pi \Gamma (2\a+k)} \right] ^{\frac{1}{2}}~.
\ee

Using the recursive relation of the Gegenbauer polynomials,  one can calculate the matrix elements of the symmetry generators in the basis \eqref{psi_n=1} and find
\be\label{<z>}
\langle \psi_l |Z| \psi_k \rangle = \sqrt{k(k-1+2\a)}\,\, \d_{l,k-1}~,\quad
\langle \psi_l| Z^* | \psi_k \rangle = \sqrt{(k+1)(k+2\a)} \,\, \d_{l,k+1}~.
\ee
These matrix elements and the spectrum $E_k=\a+k$ correspond to the well known unitary irreducible representation of the $\mathfrak{so}(2,1)$ algebra with the Casimir invariant $C=\a(\a-1)$.

Note that the parameterization $x=-\cos\sigma,\,$  $\,\s\in(0,\pi),\,$   gives to \eqref{E operator u} the form of
the Schr\"odinger  operator with  the P\"oschl-Teller potential \cite{Poeschl Teller}
\be\label{E^2 operator 2}
E^2=-\p^2_\s+\frac{\a(\a-1)}{\sin^2\s}~.
\ee

\subsection*{Eigenfunctions in $\mbox{AdS}_3$}

The symmetry generators in $\mbox{AdS}_3$ realize the decomposition $\mathfrak{so}(2,2)=\mathfrak{so}(2,1)\oplus \mathfrak{so}(2,1)$ in terms of the left and right operators
\be\label{l,r operators}
E_L=E-J_{12}~,\quad Z_L=Z_1-iZ_2~; \quad\quad E_R=E+J_{12}~,\quad Z_R^*=Z_1+iZ_2~.
\ee
This means that the triplets $(E_L, Z_L, Z_L^*)$ and $(E_R, Z_R, Z_R^*)$ both form a copy of the $\mathfrak{so}(2,1)$ algebra and the left operators commute with the right ones. Therefore, the states $\Psi_{lr}=(Z_L^*)^l\,(Z_R^*)^r\,\Psi_{E_0}$
with different pairs $(l,\,r)$ are orthogonal to each other.

In the complex coordinates on the disk $\zeta=x_1+ix_2,$ $\bar\zeta=x_1-ix_2$ the operator
\eqref{E^2 N} becomes
\be\label{E^2 3}
E^2=\zeta^2\,\p_{\zeta\zeta}^2+\bar\zeta^2\,\p_{\bar\zeta\bar\zeta}^2-(4-2\zeta\bar\zeta)\p_{\zeta\bar\zeta}^2
+2(\zeta\p_\zeta+ \bar\zeta\p_{\bar\zeta})+
\frac{1}{4}+\frac{(E_0-1)^2-\frac{1}{4}}{1-\zeta\bar \zeta}~,
\ee
and the corresponding vacuum wave function \eqref{vacuum N} is
\be\label{vaccum 3}
\Psi_{E_0}  (\zeta,\bar\zeta)\sim(1-\bar\zeta\zeta)^{\frac{E_0}{2}-\frac{1}{4}}~.
\ee
The creation operator $Z_L^*$ can be written as
\be\label{Z_l,r}
Z_L^*=\sqrt E\left(\zeta(E+1/2+\zeta\p+\bar\zeta\bar\p)-2\bar\p\right)\frac{1}{\sqrt E}~,
\ee
and $Z_R^*$ is obtained by complex conjugation of \eqref{Z_l,r}. Calculating the action of these raising operators on the eigenfunctions of energy,  one can neglect the $\sqrt E$ factors in \eqref{Z_l,r}, since they become numbers.
This remark simplifies the recursive relations between the wave functions and one finds
\be\label{lr state}
\Psi_{lr}  (\zeta,\bar\zeta)\sim ~\Psi_{E_0}  (\zeta,\bar\zeta)\,\sum_j^{\mbox{min}(l,r)}(-1)^j\frac{\Gamma(E_0  +l+r-j)}{j!\,(l-j)!\,(r-j)!}\,
\zeta^{l-j}\,\bar\zeta^{\,r-j}~,
\ee
which is an eigenfunction of the operator \eqref{E^2 3} with the eigenvalue $(E_0  +l+r)^2$.


\begin{thebibliography}{99}
\addtolength{\parskip}{-1ex}

\bibitem{B-D}
 N.~D.~Birrell and P.~C.~W.~Davies,
  \emph{``Quantum Fields In Curved Space,''}
  Cambridge, Uk: Univ. Pr. (1982) 340p.

\bibitem{fronsdal}
  C.~Fronsdal,
  \emph{``Elementary particles in a curved space. II,''}
  \href{http://dx.doi.org/10.1103/PhysRevD.10.589}{Phys.\ Rev.\  D {\bf 10}, 589 (1974)}.

\bibitem{ads-review}
   O.~Aharony, S.~S.~Gubser, J.~M.~Maldacena, H.~Ooguri and Y.~Oz,
  \emph{``Large $N$ field theories, string theory and gravity,''}
  \href{http://dx.doi.org/10.1016/S0370-1573(99)00083-6}{Phys.\ Rept.\  {\bf 323}, 183 (2000)}
  \href{http://arxiv.org/abs/hep-th/9905111}{[arXiv:hep-th/9905111]}.

\bibitem{DeWitt-R}
  B.~S.~DeWitt,
  \emph{``Point Transformations In Quantum Mechanics,''}
  \href{http://dx.doi.org/10.1103/PhysRev.85.653}{Phys.\ Rev.\  {\bf 85}, 653 (1952)}.
$\bullet$
B.~S.~DeWitt,
 \emph{``Dynamical theory in curved spaces. 1. A Review of the classical and quantum
  action principles,''}
  \href{http://dx.doi.org/10.1103/RevModPhys.29.377}{Rev.\ Mod.\ Phys.\  {\bf 29}, 377 (1957)}.

\bibitem{basti}
  F.~Bastianelli,
  \emph{``Path integrals in curved space and the worldline formalism,''}
  \href{http://arxiv.org/abs/hep-th/0508205}{arXiv:hep-th/0508205}.

\bibitem{DJ1}
    H.~Dorn and G.~Jorjadze,
  \emph{``Massless scalar particle on AdS spacetime: Hamiltonian reduction and
  quantization,''}
  \href{http://dx.doi.org/10.1088/1126-6708/2006/05/062}{JHEP {\bf 0605}, 062 (2006)}
  \href{http://arxiv.org/abs/hep-th/0508072}{[arXiv:hep-th/0508072]}.

\bibitem{B-F}
  P.~Breitenlohner and D.~Z.~Freedman,
  \emph{``Stability In Gauged Extended Supergravity,''}
  \href{http://dx.doi.org/10.1016/0003-4916(82)90116-6}{Annals Phys.\  {\bf 144}, 249 (1982)}.

\bibitem{Dorn:2005ja}
  H.~Dorn and G.~Jorjadze,
  \emph{``Oscillator quantization of the massive scalar particle dynamics on AdS
  spacetime,''}
  \href{http://dx.doi.org/10.1016/j.physletb.2005.08.059}{Phys.\ Lett.\  B {\bf 625}, 117 (2005)}
  \href{http://arxiv.org/abs/hep-th/0507031}{[arXiv:hep-th/0507031]}.

\bibitem{Avis:1977yn}
  S.~J.~Avis, C.~J.~Isham and D.~Storey,
  \emph{``Quantum Field Theory In Anti-De Sitter Space-Time,''}
  \href{http://dx.doi.org/10.1103/PhysRevD.18.3565}{Phys.\ Rev.\  D {\bf 18}, 3565 (1978)}.

\bibitem{N-W}
  T.~D.~Newton and E.~P.~Wigner,
  {\it ``Localized States For Elementary Systems,''}
  \href{http://dx.doi.org/10.1103/RevModPhys.21.400}{Rev.\ Mod.\ Phys.\  {\bf 21}, 400 (1949)}.


\bibitem{Jackiw}
   R.~Jackiw, 
  \emph{``Weyl symmetry and the Liouville theory,''}
  \href{http://dx.doi.org/10.1007/s11232-006-0090-9}{Theor.\ Math.\ Phys.\  {\bf 148}, 941  (2006)}
  \href{http://arxiv.org/abs/hep-th/0511065}{[arXiv:hep-th/0511065]}.

\bibitem{Castagnino:1989vf}
  M.~Castagnino and R.~Ferraro,
  \emph{``Hamiltonian diagonalization in foliable space-times: a Method to find the
  modes,''}
  \href{http://dx.doi.org/10.1103/PhysRevD.40.4188}{Phys.\ Rev.\  D {\bf 40}, 4188 (1989)}.

\bibitem{Wang:2009ay}
  T.~Vachaspati, D.~Stojkovic, L.~M.~Krauss,
  \emph{``Observation of incipient black holes and the information loss problem,''}
  \href{http://dx.doi.org/10.1103/PhysRevD.76.024005}{Phys.\ Rev.\  {\bf D76 } (2007)  024005}
  \href{http://arxiv.org/abs/gr-qc/0609024}{[arXiv:gr-qc/0609024]}.
\\$\bullet$
  J.~E.~Wang, E.~Greenwood and D.~Stojkovic,
  \emph{``Schr{\"o}dinger formalism, black hole horizons and singularity behavior,''}
  \href{http://dx.doi.org/10.1103/PhysRevD.80.124027}{Phys.\ Rev.\  D {\bf 80}, 124027 (2009)}
  \href{http://arxiv.org/abs/0906.3250}{[arXiv:0906.3250 [hep-th]]}.

\bibitem{Greenwood:2009gp}
  E.~Greenwood, E.~Halstead and P.~Hao,
  \emph{``Classical and Quantum Equations of Motion for a BTZ Black String in AdS
  Space,''}
  \href{http://dx.doi.org/10.1007/JHEP02(2010)044}{JHEP {\bf 1002}, 044 (2010)}
  \href{http://arxiv.org/abs/0912.1860}{[arXiv:0912.1860 [gr-qc]]}.

\bibitem{Gervais:1976ws}
  J.~L.~Gervais and A.~Jevicki,
  \emph{``Point Canonical Transformations In Path Integral,''}
  \href{http://dx.doi.org/10.1016/0550-3213(76)90422-3}{Nucl.\ Phys.\  B {\bf 110}, 93 (1976)}.

\bibitem{DeWitt}
  B.~S.~DeWitt,
  \emph{``The global approach to quantum field theory. Vol. 1''}
  Int.\ Ser.\ Monogr.\ Phys.\  {\bf 114 } (2003)  1-528.

\bibitem{Fronsdal:1975ac}
 C.~Fronsdal,
  \emph{``Elementary Particles in a Curved Space. 4. Massless
Particles,''}
  \href{http://dx.doi.org/10.1103/PhysRevD.12.3819}{Phys. \ Rev.\  {\bf D12 } (1975)  3819}

\bibitem{barata}
  N.~Yokomizo and J.~C.~A.~Barata,
  \emph{``Localizability in de Sitter space,''}
  \href{http://arxiv.org/abs/1011.0901}{arXiv:1011.0901 [hep-th]}.


\bibitem{Poeschl Teller}
G.~P{\"o}schl and E.~Teller,
\emph{``Bemerkungen zur Quantenmechanik des anharmonischen Oszillators,''}
\href{http://dx.doi.org/10.1007/BF01331132}{Z. Physik {\bf 83}, 143 (1933)}.



\end{thebibliography}
\end{document}